\documentclass{article}
\usepackage[preprint]{spconf}
\usepackage{amsmath,graphicx, url,times, enumitem}
\usepackage{color}
\usepackage{tikz}
\usepackage{pgfplots}
\usepackage{scalerel}
\usepackage{multirow}
\usepackage{makecell}
\usepackage{upgreek}
\usepackage{cite}
\usepackage{tikz-qtree}
\usepackage{caption}
\usepackage{subcaption}
\usepackage[export]{adjustbox}
 \usepackage{booktabs}
 \usetikzlibrary {shapes.multipart}
 \usetikzlibrary{shapes.geometric, arrows.meta, positioning, decorations.pathreplacing, calligraphy, fit, quotes, chains}
 \usepackage{amssymb}
\usepackage{pifont}
\newcommand{\cmark}{\ding{51}}%
\pgfplotsset{compat=1.7}
\toappear{ \makecell{\copyright\  2024 IEEE. Personal use of this material is permitted. Permission from IEEE must be obtained for all other uses, in any current or future \\media, including reprinting/republishing this material for advertising or promotional purposes, creating new collective works, for resale or \\redistribution to servers or lists, or reuse of any copyrighted component of this work in other works.}}

\title{Class-Incremental learning for multi-label audio classification}
\name{Manjunath Mulimani, Annamaria Mesaros\thanks{This work  was supported in part by Academy of Finland grant 332063 ``Teaching machines to listen". The authors wish to thank CSC-IT Centre of Science Ltd., Finland,  for providing computational resources.}}
\address{Computing Sciences, Tampere University,
Tampere, Finland}

\begin{document}
\ninept
\maketitle

\begin{abstract}
In this paper, we propose a method for class-incremental learning of potentially overlapping sounds for solving a sequence of multi-label audio classification tasks. 
We design an incremental learner that learns new classes independently of the old classes. To preserve knowledge about the old classes, we propose a cosine similarity-based distillation loss that minimizes discrepancy in the feature representations of subsequent learners, and use it along with a Kullback-Leibler divergence-based distillation loss that minimizes discrepancy in their respective outputs. 
Experiments are performed on a dataset with 50 sound classes, with an initial classification task containing 30 base classes and 4 incremental phases of 5 classes each. After each phase, the system is tested for multi-label classification with the entire set of classes learned so far. 
The proposed method obtains an average F1-score of 40.9\% over the five phases, ranging from 45.2\% in phase 0 on 30 classes, to 36.3\% in phase 4 on 50 classes. 
Average performance degradation over incremental phases is only 0.7 percentage points from the initial F1-score of 45.2\%.
\end{abstract}
\section{Introduction}
\label{sec:intro}

Incremental or continual learning is defined as training a learner for several tasks sequentially, without forgetting knowledge obtained from the preceding tasks. The data in the old tasks are not available anymore when training new tasks. 
Continuing to train a system for a new task with new data leads to what is called catastrophic forgetting\cite{parisi2019continual}--the system learns the new task but its performance on the previous task deteriorates.

In this work, our focus is on \textit{class-incremental learning} (CIL) \cite{rebuffi2017icarl}, which is one of the three identified scenarios of incremental learning as described in \cite{van2019three}. CIL uses an expanding classifier to learn classes of incrementally occurring new tasks. As such, it is different from the task-incremental learning scenario, which typically uses a separate classifier for each task \cite{li2017learning}. 
A large number of CIL studies can be found in image classification; many approaches attempt to maintain the knowledge about the previous tasks using exemplar-memory, which stores a small number of samples of the old classes for use when training new tasks\cite{lopez2017gradient, liu2021adaptive, mittal2021essentials, tiwari2022gcr}. The storage of exemplars may not be possible in some applications due to privacy and computational constraints. Other studies on CIL  use synthetic data of the old classes \cite{ye2022self}, knowledge distillation, and other regularization-based methods \cite{smith2023closer} to preserve the knowledge of previous tasks without using exemplars. 

Most research on CIL is currently available in the image processing domain, with successive tasks defined as learning new classes of objects in incremental phases. At each phase, a learner is presented with new data and new categories; after learning them, it is expected to have a good performance on both the categories learned at the current phase, as well as those learned earlier. A small number of studies attempt the same task in audio classification by learning new sound classes at incremental phases in different scenarios. Some studies apply CIL for single-class sound classification  \cite{bayram2021incremental}\cite{wang2022learning}. However, because sounds are rarely encountered in isolation, learning to classify isolated sounds is quite unrealistic when it comes to real-life audio examples. 

Incremental learning for multi-label audio classification is studied in \cite{wang2021few} using a few-shot classification setup. The method samples "pseudo" few-shot classification tasks from training data of old classes to train a classifier for new classes through meta-learning. The performance of the new classes is dependent on the frozen feature extractor which was trained using the old classes. In contrast, in a pure CIL setup, 
the entire learner (both feature extractor and classifier components) is trained by visiting each task's data exactly once. 
Recently, a CIL setup was used to solve single-class acoustic scene classification as the initial phase and multi-label sound event classification as the incremental phase  \cite{mulimani2023incremental}.    
However,  these studies were designed and tested for a single incremental audio classification task and represent only a limited view of the continual learning process that should generally consist of multiple incremental tasks over time. The effect of learning multiple consecutive tasks is largely unknown.

In this work, we propose an approach for class-incremental multi-label audio classification in multiple incremental phases. 
In this setup, new sound classes are added at incremental tasks, with multiple labels per clip possible. After each stage, when tested, the learner should assign to a test audio clip labels learned so far in any of the incremental phases (e.g. a clip may contain \textit{dog barking} and \textit{car}, with the \textit{car} class learned in task 1, and \textit{dog barking} class learned in task 3). 
This work uses the independent learning (IndL) mechanism proposed in our previous work  \cite{mulimani2023incremental}  for learning acoustic scenes and sound events.

The novelty of this work is as follows:
\begin{itemize}[noitemsep,nolistsep,leftmargin=*]
    \item We propose combining IndL with a cosine similarity-based distillation loss
    that preserves the knowledge of old sound classes by minimizing the discrepancy between features learned by successive incremental learners.    
    \item We investigate the addition of a Kullback-Leibler (KL) divergence distillation loss to the above. Cosine similarity-based distillation loss and Kullback-Leibler (KL) divergence distillation loss preserve the knowledge of old classes by minimizing the discrepancy between both features and outputs of current and previous learners 
\end{itemize}

The rest of the paper is organized as follows: Section 2 presents the notations, baselines and proposed CIL method for multi-label audio classification. Section 3 introduces the evaluation and results. Finally, conclusions are given in Section 4. 

\section{Class-Incremental Learning}
\label{sec:format}
\subsection{Tasks setup and notations}
In our class-incremental learning setup, a sequence of audio classification tasks $\tau_0, \tau_1,...,\tau_N$ is introduced to a learner over $1 + N$ time phases ($\tau_0$ in an initial phase and $\tau_1,...,\tau_N$ in $N$ incremental phases).
Each task  $\tau_i$ is composed of input data $\mathcal{X}_i=\{(\mathbf{x}_j^{i}, \mathbf{y}_j^{i})|1\leq j\leq m\}$, where $\mathbf{x}^{i}$ is input features and $\mathbf{y}^{i}\in \{0, 1\}^{|C|}$ is the multi-hot vectors encoding the ground truth. $|C|=|C_{old}|+|C_{new}|$ denotes the number of sound classes from previous tasks ($|C_{old}|$) and new sound classes in the current task ($|C_{new}|$).

A learner $\mathcal{P}^i$ that learns task $\tau_i$ is a deep network that includes a feature extractor $\mathcal{F}_\mathbf{\theta}^i$, with parameters $\theta$ and a classifier $\mathcal{H}_\phi^i$, with parameters $\phi$. 
Output logits of the network on a given input $\mathbf{x}$ are obtained by $\mathbf{o}(\mathbf{x}) = \mathcal{H}_\phi^i(\mathcal{F}_\theta^i(\mathbf{x}))$. 

\subsection{Baselines}
We construct a few standard baselines under the CIL setup:  
(1) \textit{Fine-tuning (FT)}: a learner is fine-tuned on the new classes at each incremental phase  without taking any measures to mitigate catastrophic forgetting; (2) \textit{Feature extraction (FE)}: the feature extractor component $\mathcal{F}_\mathbf{\theta}^i$ is frozen after learning the classes of the initial phase. In an incremental phase, classification weight vectors of the old classes in $\mathcal{H}_\phi^i$ are frozen, and only the weight vectors of the new classes are trained; 
(3) \textit{Audio tagging (AT)}: a single non-incremental learner is trained from scratch at each phase on data of all classes from previous and current tasks.

\subsection{Class-incremental learning method}
Similar to standard multi-label audio classification systems, an incremental learner learns to solve the base task $\tau_0$ on data $\mathcal{X}_0$ using binary cross-entropy loss ($\mathcal{L}^{BCE}$). $\mathcal{L}^{BCE}$ is computed in both initial and incremental phases using sigmoid $\sigma$ over logits of the current task classes:
 \begin{multline}
     \mathcal{L}^{BCE} = -\sum_{k=|C_{old}|}^{|C|}\mathbf{y}_k^{{i}}\cdot\log(\sigma(\mathbf{o}_k)) \\ +(1-\mathbf{y}_k^{{i}})\cdot\log(1-\sigma(\mathbf{o}_k)),
      \label{eq1}
 \end{multline}
 
\begin{figure}[!tbp]
  \centering
  \includegraphics[width=\linewidth]{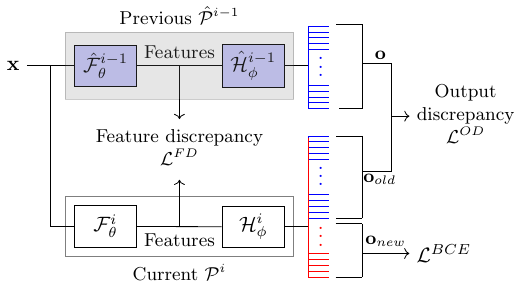}  
  \caption{An overview of the proposed CIL approach in an incremental time phase $i$. Three losses: $\mathcal{L}^{BCE}$, $\mathcal{L}^{OD}$ and $\mathcal{L}^{FD}$ are used to train current learner $\mathcal{P}^{{i}}$. $\mathcal{L}^{OD}$ and $\mathcal{L}^{FD}$ minimize the output and feature discrepancy between $\mathcal{P}^{{i}}$ and frozen $\hat{\mathcal{P}}^{i-1}$ learners to preserve the knowledge of previous classes. $\mathcal{L}^{BCE}$ is computed independently over logits $\mathbf{o}_{new}$ to learn new classes.}
     \label{fig:ODFD}
     \vspace{-10pt}
\end{figure}

An overview of the proposed CIL approach in an incremental time phase $i$ is given in  Fig.~\ref{fig:ODFD}. 
We obtain a new learner $\mathcal{P}^{i}$ by adding new output units to the classifier $\mathcal{H}_\phi^{{i-1}}$ of previous learner $\mathcal{P}^{i-1}$, to learn the new classes of the current task $\tau_i$. Specifically, $\mathcal{P}^{i}$ is $\mathcal{P}^{i-1}$ with additional output units in the classifier. In each incremental phase, we only extend the classifier $\mathcal{H}_\phi^{{i-1}}$ of $\mathcal{P}^{i-1}$ to get $\mathcal{P}^{i}$. 
Therefore, output logits of $\mathcal{P}^{{i}}$ comprise  $\mathbf{o}=\{\mathbf{o}_{old}, \mathbf{o}_{new}\}$. $\mathbf{o}_{old}$ and $\mathbf{o}_{new}$ denote the logits of $C_{old}$ and $C_{new}$ classes respectively.
The entire new $\mathcal{P}^{{i}}$ ($\mathcal{F}_\mathbf{\theta}^i$ and $\mathcal{H}_\phi^{{i}}$) is trained on current data $\mathcal{X}_{i}$ to solve task $\tau_{i}$, in the absence of the previous tasks data $\mathcal{X}_{0:i-1}$. The training is controlled using three different losses as explained later.

In a conventional training process, the data imbalance between previous tasks $\tau_{0:i-1}$ (no data) and a current task $\tau_{i}$ makes the learner's predictions biased to focus on the $C_{new}$ classes and catastrophically forget the $C_{old}$ classes.
To reduce this catastrophic forgetting, we use the independent learning (IndL) mechanism introduced in \cite{mulimani2023incremental} to allow learning of $\mathbf{o}_{new}$ logits independently from $\mathbf{o}_{old}$ logits.  To achieve this, we compute $\mathcal{L}^{BCE}$ using sigmoid $\sigma$ only over $\mathbf{o}_{new}$ logits as per Eq. \ref{eq1}.
The  $\mathbf{o}_{old}$ logits are processed separately.

To preserve the previous knowledge, we propose a combination of two distillation losses: namely, output discrepancy loss $\mathcal{L}^{OD}$ and feature discrepancy loss $\mathcal{L}^{FD}$. $\mathcal{L}^{OD}$ minimizes the output discrepancy between $\mathcal{P}^{{i}}$ and ${\mathcal{P}}^{i-1}$ on the output logits of the old classes using Kullback-Leibler divergence ($\mathcal{D}_{KL}$). On the other hand, $\mathcal{L}^{FD}$ minimizes feature discrepancy between $\mathcal{P}^{{i}}$ and ${\mathcal{P}}^{i-1}$  on the normalized output features using cosine similarity. 
Their role in learning is illustrated in Fig.~\ref{fig:ODFD}: in essence, $\mathcal{L}^{OD}$ is applied at the output of the classifier $\mathcal{H}_\phi^{{i}}$, while $\mathcal{L}^{FD}$ is applied at the output of the feature extractor $\mathcal{F}_\mathbf{\theta}^i$, with the aim to make the current learner mimic the previous learner's behavior on old classes in the absence of data examples for those classes.  
For $\mathcal{L}^{OD}$, the output logits $\mathbf{o}_{old}$ of the current learner ${\mathcal{P}}^{i}$ are compared with the output logits of the frozen previous learner $\hat{\mathcal{P}}^{i-1}$:
\begin{equation}
    \mathcal{L}^{OD} =\mathcal{D}_{KL}(\pi(\hat{\mathcal{P}}^{i-1}(\mathbf{x}))||\pi(\mathbf{o}_{old})),
    \label{eq2}
\end{equation}
where $\pi(\mathbf{u})=\mathbf{u}_i^{1/\Delta}/\Sigma_j\mathbf{u}_j^{1/\Delta}$ is a rescaling function. The value of $\Delta$ is set to higher than 1  to increase the weights of smaller values\cite{hou2019learning}.
$\mathcal{L}^{FD}$ is computed by comparing the features $\mathbf{v}=\mathcal{F}_\mathbf{\theta}^{i}(\mathbf{x})$ from the current learner ${\mathcal{P}}^{i}$  with the features $\hat{\mathbf{v}}=\hat{\mathcal{F}}_\mathbf{\theta}^{i-1}(\mathbf{x})$ from  $\hat{\mathcal{P}}^{i-1}$:
\begin{equation}
    \mathcal{L}^{FD} = 1-cos(norm(\hat{\mathbf{v}}), norm(\mathbf{v})),
    \label{eq3}
\end{equation}
where $norm$ and $cos$ denote the L2-normalization and cosine similarity respectively. The incremental learner is trained using the combined losses as: 
\begin{equation}
    \mathcal{L} = \mathcal{L}^{BCE}+\mathcal{L}^{FD}+\lambda\mathcal{L}^{OD},
    \label{eq6}
\end{equation}
where the weight $\lambda=\Omega\sqrt{|C|/|C_{new}|}$ is set adaptively as proposed in \cite{hou2019learning}, \cite{kang2022class}. $\Omega$ is a fixed constant and $\lambda$ increases as the number of incremental phases increases, to retain the knowledge of the increasing number of old classes in the classifier. Hereafter, our proposed approach is referred to as IODFD (I: IndL, OD: $\mathcal{L}^{OD}$, FD: $\mathcal{L}^{FD}$). We also compare with IOD (I: IndL, OD: $\mathcal{L}^{OD}$) and IFD (I: IndL, FD: $\mathcal{L}^{FD}$) in \ref{res} to investigate the role of each in the combination. 

\begin{figure}[!tbp]
  \centering
  \subfloat{\includegraphics[width=\linewidth, height=3cm]{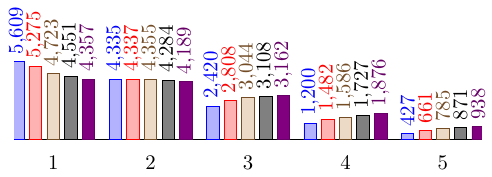}}\\
    \subfloat{\includegraphics[width=\linewidth, height=3cm]{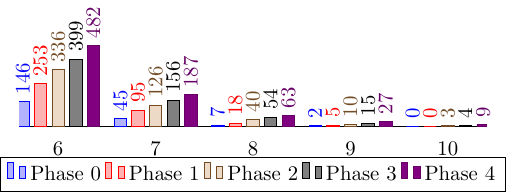}}
     \caption{Number of labels per file in the evaluation set, as the incremental learning of new classes progresses up to 50 classes.}
    \label{fig:poly}
\vspace{-10pt}
\end{figure}

\section{Evaluation and Results}
\label{sec:evr}

\subsection{Dataset and training setup}

Experiments are performed on a subset of the temporally strong Audioset; we use the labels as weak labels only, but benefit of the higher quality labels than the original Audioset weak labels, due to the annotation method \cite{hershey2021benefit}. 
We selected the 50 largest classes for our experiments.
The initial phase 0 consists of 30 classes, selected randomly; each subsequent incremental phase adds 5 classes to the system, selected randomly from the remaining 20 classes. We maintain the same order for all experiments. 

In phase $i$, all clips labeled with classes to be learned at this phase are used as training examples. The clips may contain classes from other, past or future, incremental phases, which we ignore; we use only information about the current target classes. 
In practice, 
there is a possibility that the same clip is used in different incremental phases with different subsets of its labels; we estimate that in our experiments there is about 10-12\% overlap. A separate experiment removing this overlap at all stages was performed and showed that this overlap did not influence the final performance.
For AT at phase $i$, all clips in the development set labeled with classes of previous and current phases of the equivalent incremental learner are used as training examples, and the system is trained from scratch.
Note that classes in all phases are severely imbalanced in terms of size.

After each incremental phase $i$, we evaluate multi-label classification with all classes learned so far (phases 0 to $i$); classes to be learned in upcoming phases are not evaluated. The multi-class classification problem has an increasing number of sound classes to be recognized after each incremental phase; additionally, for each clip there are more concurrent and potentially overlapping sounds to be recognized. Figure ~\ref{fig:poly} shows the average number of labels per clip in the evaluation dataset at each phase: after phase 0 there are 427 clips with 5 labels (of the 30 classes learned at phase 0), while after phase 4 there are 938 clips with 5 labels (of the 50 classes learned in total).

\begin{figure*}[!tbp]
  \centering
  \subfloat{\includegraphics[width=5.8cm, height=4.9cm]{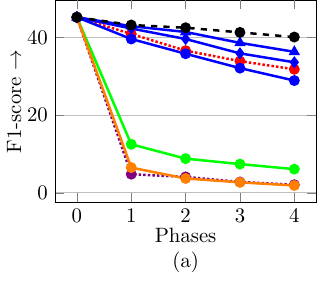}}
  \hfill
  \subfloat{\includegraphics[width=5.8cm, height=4.9cm]{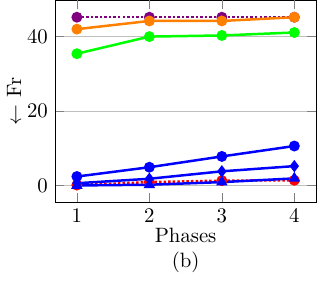}}
  \hfill
  \subfloat{\includegraphics[width=5.8cm, height=4.9cm]{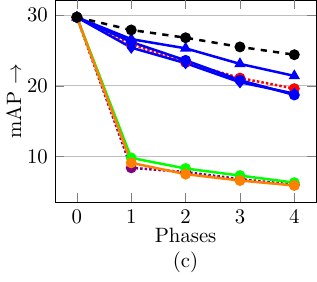}}\\
  \subfloat{\includegraphics[width=11cm]{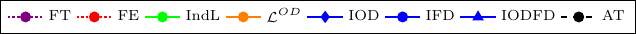}}
 \caption{Comparison of F1-score (a), Fr (b) and mAP (c) of approaches.  Approaches are the non-incremental audio tagging (AT), fine-tuning (FT), feature extraction (FE), IndL (without distillation losses), $\mathcal{L}^{OD}$ (without IndL),   IOD \cite{mulimani2023incremental}, proposed IFD and IODFD.}
        \label{Fig:CMP}
     \vspace{-10pt}
\end{figure*}
 
\subsection{Implementation details and evaluation metrics}

We use PANNs CNN14 \cite{kong2020panns} as a feature extractor $\mathcal{F}_\mathbf{\theta}^{i}$. 
The global pooling is applied to the last convolutional layer to get a fixed-length input feature vector to the classifier $\mathcal{H}_\phi^{i}$. The number of output units in $\mathcal{H}_\phi^{i}$ is equal to the total number of classes from previous and current phases. For a fair comparison, all the systems reported in this work use the same network architecture. We compute log mel spectrogram of size $1001\times 64$ from a 10-second audio recording as input to the CNN14, using the configuration provided in \cite{kong2020panns}.

The entire learner network is trained using the SGD optimizer \cite{loshchilov2017sgdr} with a momentum of 0.9 and a mini-batch size of 32 for 120 epochs. 
Reducing the learning rate when fine-tuning the learner on new data is a standard method in transfer learning. We reduce the learning rate in incremental phases to improve the knowledge transfer between the base and incremental phases as done in \cite{mittal2021essentials}, \cite{mulimani2023incremental}.
The learning rates for the initial phase 0 and the incremental phases are set to \{0.01 and 0.001\}. 
This combination was chosen based on a number of preliminary experiments performed with different learning rates at different phases, which we omit for brevity. We use a mini-batch balanced sampling strategy as per the recommendation of \cite{kong2020panns} to balance the classes in each phase.
CosineAnnealingLR \cite{loshchilov2017sgdr} scheduler is used to update the optimizer in every epoch. The $\Omega$ and $\Delta$ are empirically set to 2.  

The performance of the learner is evaluated using the class-wise average (macro) F1-score with a threshold of 0.5 for positive output, and mAP (mean average precision).
We also measure Forgetting (Fr) as the difference between the F1-score of base and current incremental learners over the initial 30 classes, i.e. the performance drop on the base classes after learning each consecutive set of new classes. A lower Fr is better, showing that the system maintains knowledge about the classes learned in the base task throughout the ulterior training phases.
\begin{table}[]
 %\scriptsize
     \centering
   \begin{tabular}{l|c|c|c|c|c|c}
 \toprule
 Method & IndL & $\mathcal{L}^{OD}$ & $\mathcal{L}^{FD}$ & F1 $\uparrow$& Fr $\downarrow$& mAP $\uparrow$\\
 \cline{1-7}
 FT &-&-&-&11.8 & 45.2 &11.8\\
 %\hline
 FE &-&-&-& 37.7 & 0.9 &23.9\\
 \hline
IndL &\cmark &- &- & 16.0 & 39.2 &12.3\\
$\mathcal{L}^{OD}$ &-&\cmark &-& 12.0 & 43.9 &11.8\\
 \hline
 IOD\cite{mulimani2023incremental} &\cmark&\cmark&-& 39.4 & 2.9 &23.6\\
 %\hline
  IFD &\cmark&-&\cmark& 36.4 & 6.4 &23.8\\
 %\hline
 IODFD &\cmark&\cmark&\cmark& \textbf{40.9} & \textbf{0.7} &\textbf{25.3}\\
  \hline
AT &-&-&-& {42.4} & {-} &{26.8}\\
 \bottomrule
 \end{tabular}
     \caption{Average F1-score and mAP over the five phases (initial + incremental). Average Fr over four incremental phases.}
     \label{tab:Avg}
     \end{table}
\begin{table}[]
     \centering
   \begin{tabular}{l|c|c|c}
 \toprule
 Method & $C_{old}$ (0-44) & $C_{new}$ (45-49) & Overall (0-49)\\
 \cline{1-4}
 FT & 0.0 & 20.6 &2.1\\
% \hline
 FE & 34.2 & 10.6 &31.8\\
 \hline
 IOD\cite{mulimani2023incremental} & 34.8 & 22.2 &33.6\\
% \hline
 IODFD & \textbf{37.8} & \textbf{23.1} &\textbf{36.3}\\
 \bottomrule
  \end{tabular}
     \caption{F1-scores computed separately for old and new classes after incremental phase 4; overall F1-score over all 50 classes.}
     \label{tab:ATE}
     \vspace{-10pt}
     \end{table}
\begin{figure}[htbp]
  \centering
  \subfloat{\includegraphics[width=4.0cm, height=4.21cm]{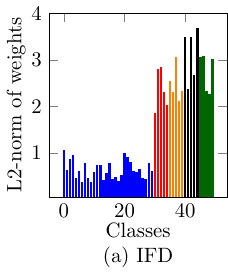}}
  \hfill
    \subfloat{\includegraphics[width=4.0cm, height=4.1cm]{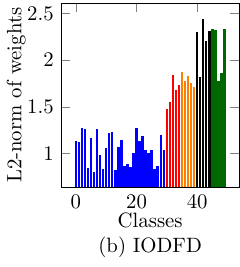}}
     \caption{L2-norm of the classification weight vectors for all the classes at the end of incremental learning (phase 4).}
    \label{fig:weight}
    \vspace{-10pt}
\end{figure}

\subsection{Results}
\label{res}

The learner trained from scratch on the 30 base classes achieves the F1-score of 45.2\% and mAP of 29.7\%. This learner is further used to learn tasks in four incremental phases using different approaches (except AT). 
The F1-score, Fr, and mAP of these different approaches are compared in Fig.~\ref{Fig:CMP}. 
The average performance of all approaches over the five different tasks (phases 0 to 4) is presented in Table \ref{tab:Avg}. We also detail the performance on the old and new classes at the fourth incremental phase in Table \ref{tab:ATE}.
%Table \ref{tab:ATE} demonstrates the performance of different approaches on old and new classes at incremental phase 4. 

As expected, the FT approach learns to classify new sound classes, but forgets all information about the old sound classes including the 30 base classes. As seen in Table \ref{tab:ATE} (row 1), the F1-score on new classes at phase 4 is 20.6\%, but the F1 score on all the previous classes is zero; this is also reflected in Fr being equal to the F1-score from phase 0 in Table \ref{tab:Avg}. The immediate effect of learning new information can be seen also in Fig.~\ref{Fig:CMP}b: overall F1-score or mAP of FT is significantly lower in each incremental phase. On the other hand, FE learns only new classes at each phase with a fixed feature representation of the base classes. This allows FE to achieve superior results on base classes, with the least Fr (see row 2 of Table \ref{tab:Avg} or Fig.~\ref{Fig:CMP}b). In turn, FE shows poor performance on learning new classes at each incremental phase; for instance, F1 score on new classes at phase 4 (Table \ref{tab:ATE}, row 2) is much lower than for FT. Specifically, FT exhibits high plasticity, i.e. the learner significantly forgets the old sound classes. On the flip side, FE shows high stability, i.e. fails to adapt to new classes. The difference them illustrates the stability-plasticity dilemma between old and new classes. 

We investigate two more methods to show the performance of the learner either without IndL  or only with IndL ($\mathcal{L}^{OD}$ and IndL in Fig.~\ref{Fig:CMP}); both methods suffer from high plasticity, similar to FT. The learner without IndL is trained using Eq. (\ref{eq1}) over all logits $\mathbf{o}$ (both $\mathbf{o}^{old}, \mathbf{o}^{new}$) on current data $\mathcal{X}_{i}$.  This makes the learner easily forget the old sound event classes because it sees no examples of them.
IndL allows learning in incremental phases from only the classifier weights of the new classes, with a common feature extractor on current data $\mathcal{X}_{i}$. However, because the feature extractor continues to be trained with the new data,  the learner with IndL alone fails to retain the previous classes' knowledge, as observed in Fig.~\ref{Fig:CMP}.  To mitigate this, $\mathcal{L}^{OD}$ penalizes the change in output logits of old classes compared to the previous learner and effectively preserves the knowledge of old classes.
Hence, a combination of IndL and $\mathcal{L}^{OD}$ (IOD) achieves less forgetting than IndL or $\mathcal{L}^{OD}$ alone.  
While mAP and Fr of FE are better than those of IOD (see Fig.~\ref{Fig:CMP}b and ~\ref{Fig:CMP}c or Table \ref{tab:Avg}), the performance of IOD on new classes is better (Table \ref{tab:ATE}, row 3). 

The proposed IFD makes the current learner mimic the previous learners' knowledge of old classes by forcing it to produce similar feature representations. IFD reduces the feature drift between current and previous learners on old classes but fails to reduce the bias in the classifier. To analyse the effect of IFD on the classifier, we show the magnitudes of weight vectors of base and incremental classes in $\mathcal{H}_\phi^i$ after phase 4 in Fig.~\ref{fig:weight}a. It can be observed that the magnitudes of the weight vectors of classes from incremental phases are much higher than that of the classes from the initial phase due to imbalanced data between tasks. It makes IFD results biased in favor of new classes, gradually forgetting the base classes, an effect also shown by the increasing Fr in Table \ref{tab:Avg} and Fig.~\ref{Fig:CMP}b.

A combination of $\mathcal{L}^{OD}$, $\mathcal{L}^{FD}$ and IndL (IODFD) reduces the discrepancy between current and previous learners in both feature space and prediction space. The addition of $\mathcal{L}^{OD}$ to $\mathcal{L}^{FD}$ brings the magnitude of sound classes in incremental phases close to base classes  (see Fig.~\ref{fig:weight}b), enabling the classifier to detect both base and incremental classes.  The reasonably higher magnitudes of new classes also help to detect minority classes with few samples effectively in incremental phases without affecting the performance of the old classes. As a result, the proposed IODFD achieves significantly better performance on both old and new classes, with an average F1 score of 40.9\%, mAP of 25.3\% and least Fr of 0.7 percentage points over the five tasks (see Table \ref{tab:Avg}). IODFD outperforms all other methods on the classification of both old and new classes 
and its performance is close to AT, indicating that IODFD is plastic enough to learn new sound classes and stable enough to retain the knowledge of old classes. Note that mAP of CNN14 on the selected 30 base classes in our experiments is 29.7\%  whereas CNN14 in \cite{kong2020panns} is trained on the complete weakly-labeled Audioset with different augmentation techniques, achieving mAP of 43.1\%.

\section{Conclusion}
In this paper, we presented an incremental learner to solve multi-label audio classification tasks sequentially. The results show that the proposed IODFD performs significantly better than all other investigated methods to solve imbalanced sequential sound classification tasks. Either IOD or IFD alone failed to balance plasticity and stability, resulting in lower performance. 
Future work includes investigating the role of exemplars in CIL; another research direction is incremental learning for sound event detection, in which there is also temporal information about overlapping sounds.

\bibliographystyle{IEEEbib}
\bibliography{references}

\begin{thebibliography}{10}

\bibitem{parisi2019continual}
German~I Parisi, Ronald Kemker, Jose~L Part, Christopher Kanan, and Stefan Wermter,
\newblock ``Continual lifelong learning with neural networks: A review,''
\newblock {\em Neural networks}, vol. 113, pp. 54--71, 2019.

\bibitem{rebuffi2017icarl}
Sylvestre-Alvise Rebuffi, Alexander Kolesnikov, Georg Sperl, and Christoph~H Lampert,
\newblock ``icarl: Incremental classifier and representation learning,''
\newblock in {\em Proceedings of the IEEE conference on Computer Vision and Pattern Recognition}, 2017, pp. 2001--2010.

\bibitem{van2019three}
Gido~M Van~de Ven and Andreas~S Tolias,
\newblock ``Three scenarios for continual learning,''
\newblock in {\em Continual Learning Workshop NeurIPS}, 2018.

\bibitem{li2017learning}
Zhizhong Li and Derek Hoiem,
\newblock ``Learning without forgetting,''
\newblock {\em IEEE Transactions on pattern analysis and machine intelligence}, vol. 40, no. 12, pp. 2935--2947, 2017.

\bibitem{lopez2017gradient}
David Lopez-Paz and Marc'Aurelio Ranzato,
\newblock ``Gradient episodic memory for continual learning,''
\newblock {\em Advances in neural information processing systems}, vol. 30, 2017.

\bibitem{liu2021adaptive}
Yaoyao Liu, Bernt Schiele, and Qianru Sun,
\newblock ``Adaptive aggregation networks for class-incremental learning,''
\newblock in {\em Proceedings of the IEEE/CVF Conference on Computer Vision and Pattern Recognition}, 2021, pp. 2544--2553.

\bibitem{mittal2021essentials}
Sudhanshu Mittal, Silvio Galesso, and Thomas Brox,
\newblock ``Essentials for class incremental learning,''
\newblock in {\em Proceedings of the IEEE/CVF Conference on Computer Vision and Pattern Recognition}, 2021, pp. 3513--3522.

\bibitem{tiwari2022gcr}
Rishabh Tiwari, Krishnateja Killamsetty, Rishabh Iyer, and Pradeep Shenoy,
\newblock ``Gcr: Gradient coreset based replay buffer selection for continual learning,''
\newblock in {\em Proceedings of the IEEE/CVF Conference on Computer Vision and Pattern Recognition}, 2022, pp. 99--108.

\bibitem{ye2022self}
Fanfan Ye, Liang Ma, Qiaoyong Zhong, Di~Xie, and Shiliang Pu,
\newblock ``Self-distilled knowledge delegator for exemplar-free class incremental learning,''
\newblock in {\em International Joint Conference on Neural Networks (IJCNN)}. IEEE, 2022, pp. 1--8.

\bibitem{smith2023closer}
James~Seale Smith, Junjiao Tian, Shaunak Halbe, Yen-Chang Hsu, and Zsolt Kira,
\newblock ``A closer look at rehearsal-free continual learning,''
\newblock in {\em Proceedings of the IEEE/CVF Conference on Computer Vision and Pattern Recognition}, 2023, pp. 2409--2419.

\bibitem{bayram2021incremental}
Bar{\i}{\c{s}} Bayram and G{\"o}khan {\.I}nce,
\newblock ``An incremental class-learning approach with acoustic novelty detection for acoustic event recognition,''
\newblock {\em Sensors}, vol. 21, no. 19, pp. 6622, 2021.

\bibitem{wang2022learning}
Zhepei Wang, Cem Subakan, Xilin Jiang, Junkai Wu, Efthymios Tzinis, Mirco Ravanelli, and Paris Smaragdis,
\newblock ``Learning representations for new sound classes with continual self-supervised learning,''
\newblock {\em IEEE Signal Processing Letters}, vol. 29, pp. 2607--2611, 2022.

\bibitem{wang2021few}
Yu~Wang, Nicholas~J Bryan, Mark Cartwright, Juan~Pablo Bello, and Justin Salamon,
\newblock ``Few-shot continual learning for audio classification,''
\newblock in {\em IEEE International Conference on Acoustics, Speech and Signal Processing (ICASSP)}. IEEE, 2021, pp. 321--325.

\bibitem{mulimani2023incremental}
Manjunath Mulimani and Annamaria Mesaros,
\newblock ``Incremental learning of acoustic scenes and sound events,''
\newblock in {\em Proceedings of the 8th Workshop on Detection and Classification of Acoustic Scenes and Events (DCASE)}, 2023, pp. 141--145.

\bibitem{hou2019learning}
Saihui Hou, Xinyu Pan, Chen~Change Loy, Zilei Wang, and Dahua Lin,
\newblock ``Learning a unified classifier incrementally via rebalancing,''
\newblock in {\em Proceedings of the IEEE/CVF Conference on Computer Vision and Pattern Recognition}, 2019, pp. 831--839.

\bibitem{kang2022class}
Minsoo Kang, Jaeyoo Park, and Bohyung Han,
\newblock ``Class-incremental learning by knowledge distillation with adaptive feature consolidation,''
\newblock in {\em Proceedings of the IEEE/CVF conference on computer vision and pattern recognition}, 2022, pp. 16071--16080.

\bibitem{hershey2021benefit}
Shawn Hershey, Daniel~PW Ellis, Eduardo Fonseca, Aren Jansen, Caroline Liu, R~Channing Moore, and Manoj Plakal,
\newblock ``The benefit of temporally-strong labels in audio event classification,''
\newblock in {\em IEEE International Conference on Acoustics, Speech and Signal Processing (ICASSP)}. IEEE, 2021, pp. 366--370.

\bibitem{kong2020panns}
Qiuqiang Kong, Yin Cao, Turab Iqbal, Yuxuan Wang, Wenwu Wang, and Mark~D Plumbley,
\newblock ``{PANN}s: Large-scale pretrained audio neural networks for audio pattern recognition,''
\newblock {\em IEEE/ACM Transactions on Audio, Speech, and Language Processing}, vol. 28, pp. 2880--2894, 2020.

\bibitem{loshchilov2017sgdr}
Ilya Loshchilov and Frank Hutter,
\newblock ``{SGDR}: Stochastic gradient descent with warm restarts,''
\newblock in {\em International Conference on Learning Representations (ICLR)}, 2017.

\end{thebibliography}

\end{document}